\documentclass[12pt,preprint]{aastex}
\renewcommand{\cite}{\citealp}
\def\ltsim{\, {}^<_\sim \,}
\def\vmi{\hbox{\it V--I\/}}
\def\bmv{\hbox{\it B--V\/}}
\def\bmi{\hbox{\it B--I\/}}
\def\ngc#1{NGC$\,$#1}

\newcommand{\dydz}{$\Delta {\rm Y} / \Delta {\rm Z}\;$}

\newcommand{\lsun}{log $L/L_{\odot}\,$}

\newcommand{\feh}{[Fe/H]~}

\shorttitle{On a new parameter to estimate the He content in old stellar systems}
\shortauthors{Troisi et al.}

\begin{document}

\title{On a new parameter to estimate the helium content in old stellar systems\altaffilmark{1}}

\author{
F.\ Troisi\altaffilmark{2},  
G.\ Bono\altaffilmark{2,3},
P.\ B.\ Stetson\altaffilmark{4,14,15,16},
A.\ Pietrinferni\altaffilmark{5},
A.\ Weiss\altaffilmark{6},
M.\ Fabrizio\altaffilmark{2},
I.\ Ferraro\altaffilmark{3},
A.\ Di Cecco\altaffilmark{7},
G.\ Iannicola\altaffilmark{3},
R.\ Buonanno\altaffilmark{2,7},
A.\ Calamida\altaffilmark{3},
F.\ Caputo\altaffilmark{2,3},
C.\ E.\ Corsi\altaffilmark{3},
M.\ Dall'Ora\altaffilmark{8},
A.\ Kunder\altaffilmark{9},
M.\ Monelli\altaffilmark{10},
M.\ Nonino\altaffilmark{11},
A.\ M. Piersimoni\altaffilmark{5}, 
L.\ Pulone\altaffilmark{3},
M.\ Romaniello\altaffilmark{12}, 
A.\ R.\ Walker\altaffilmark{9}, and 
M.\ Zoccali\altaffilmark{13}
}   

\altaffiltext{1}{This paper makes use of data obtained from the ESO/ST-ECF Science Archive 
Facility; from the Isaac Newton Group Archive which is maintained as part of the CASU Astronomical 
Data Centre at the Institute of Astronomy, Cambridge; and from the Canadian Astronomy 
Data Centre operated by the National Research Council of Canada with the support of the 
Canadian Space Agency.} 
\altaffiltext{2}{Dipartimento di Fisica, Universit\`a di Roma Tor Vergata, via della
Ricerca Scientifica 1, 00133 Rome, Italy; licia.troisi@roma2.infn.it}
\altaffiltext{3}{INAF--Osservatorio Astronomico di Roma, via Frascati 33, 00040 Monte Porzio Catone, Rome, Italy}
\altaffiltext{4}{Dominion Astrophysical Observatory, Herzberg Institute of Astrophysics, National Research Council, 5071 West Saanich Road, Victoria, BC V9E 2E7, Canada}
\altaffiltext{5}{INAF-Osservatorio Astronomico di Collurania, Via M. Maggini, 64100 Teramo, Italy}
\altaffiltext{6}{Max-Planck-Institut fur Astrophysik, Karl-Schwarzschild-Str. 1, 85748 Garching, Germany}
\altaffiltext{7}{Agenzia Spaziale Italiana--Science Data Center, ASDC c/o ESRIN, via G. Galilei, 00044 Frascati, Italy}
\altaffiltext{8}{INAF--Osservatorio Astronomico di Capodimonte, via Moiariello 16, 80131 Napoli, Italy}
\altaffiltext{9}{Cerro Tololo Inter-American Observatory, National Optical Astronomy Observatory, Casilla 603, La Serena, Chile}
\altaffiltext{10}{Instituto de Astrof\'isica de Canarias, Calle Via Lactea, E38200 La Laguna, Tenerife, Spain}
\altaffiltext{11}{INAF--Osservatorio Astronomico di Trieste, via G.B. Tiepolo 11, 40131 Trieste, Italy}
\altaffiltext{12}{European Southern Observatory, Karl-Schwarzschild-Str. 2, 85748 Garching bei Munchen, Germany}
\altaffiltext{13}{Departamento Astronomía y Astrofísica, Pontificia Universidad Católica de Chile, 
Av. Vicuna Mackenna 4860, Santiago, Chile }
\altaffiltext{14}{Visiting Astronomer, Kitt Peak National Observatory, National Optical Astronomy
Observatory, which is operated by the Association of Universities for Research in Astronomy (AURA)
under cooperative agreement with the National Science Foundation.}
\altaffiltext{15}{Visiting astronomer, Cerro Tololo Inter-American Observatory, National Optical 
Astronomy Observatory, which are operated by the Association of Universities for Research in Astronomy, 
under contract with the National Science Foundation.}
\altaffiltext{16}{Visiting Astronomer, Canada-France-Hawaii Telescope operated by the
National Research Council of Canada, the Centre National de la Recherche
Scientifique de France and the University of Hawaii.}

%%%%%%%%%%%%%%%%%%%%%%%%%%%%%%%%%%%%%%%%%%%%%%%%%%%%%%%%%%%%%%%%%%%%%%%%%%%%%%%
\begin{abstract}
We introduce a new parameter $\Delta \xi$---the difference in magnitude between the 
red giant branch (RGB) bump and the  point on the main sequence (MS) at the same color 
as the bump (which we call the ``benchmark'')---to estimate the helium content in old stellar systems. 
The $\Delta\xi$ parameter is a helium indicator since an increase in helium makes, 
at fixed age and iron abundance, the RGB bump brighter and the MS benchmark 
fainter. Moreover, its sensitivity to helium is linear over the entire metallicity range. 
$\Delta \xi$ is also minimally affected by changes of a few Gyr in cluster 
age, by uncertainties in the photometric zero-point, by the amount of reddening,
or by the effects of evolution on the horizontal branch. The two main drawbacks of the 
$\Delta \xi$ parameter include the need for precise and large photometric data sets   
from the RGB bump down to the MS benchmark, and a strong dependence of the 
$\Delta \hbox{\rm Y} / \Delta \xi$ slope on metallicity.  
To provide an empirical basis for the $\Delta \xi$ parameter we selected almost two 
dozen relatively bright Galactic Globular Clusters (GGCs) with low foreground reddening, 
and a broad range of iron abundance ($-2.45\le$[Fe/H]$\le-0.70$ dex). Moreover, the selected GGCs 
have precise, relatively deep, and homogeneous multi-band ({\it BVI\/}) photometry.
We found that the observed $\Delta \xi$ parameters and those predicted from
$\alpha$-enhanced evolutionary models agree reasonably well if we
assume a primordial helium content of Y=0.20 (abundance by mass). The discrepancy in the 
photometric $B$ band becomes of the order of 4$\sigma$ ($\Delta B=0.20$~mag) only in the 
metal-poor regime. Comparison with evolutionary prescriptions 
based on a canonical primordial helium content (Y=0.245, $\Delta$Y/$\Delta$Z [helium-to-metal 
enrichment ratio]=1.4) indicates that the observed $\Delta \xi$ values are systematically 
smaller than predicted. This discrepancy ranges from  5$\sigma$
($\Delta B=0.26$~mag) in the metal-rich regime to 10$\sigma$
($\Delta B=0.51$~mag) in the metal-poor regime. The outcome is the
same if predicted $\Delta \xi$ parameters are based on
evolutionary models with CNO enhancements in addition to $\alpha$
enhancements.  The discrepancy becomes even larger if we extend
the comparison to He-enhanced models. These findings support
previous results (Meissner \& Weiss 2006, \aap, 456, 1085;
Di~Cecco et al.\ 2010, \apj, 712, 527; Cassisi et al.\ 2011, \aap,
527, A59) suggesting that current stellar evolutionary models
overestimate the luminosity of the RGB bump.      
We also found that including envelope overshooting can eliminate the discrepancy, as 
originally suggested by Alongi et al.\ (1993, \aaps, 97, 851);
atomic diffusion and mass loss play smaller roles. 
The $\Delta \xi$ parameter of GGCs, in spite of the possible 
limitations concerning the input physics of current evolutionary models, provides 
an independent detection of pre-stellar helium at least at the 5$\sigma$ level.
\end{abstract}


%%%%%%%%%%%%%%%%%%%%%%%%%%%%%%%%%%%%%%%%%%%%%%%%%%%%%%%%%%%%%%%%%%%%%%%%%%%%%%%%%%%%%
\section{Introduction} 

The notion of possible variations in helium abundance between different
Globular Clusters (GCs) dates back more than forty years \citep{vand67}, in
connection with the second parameter phenomenon \citep{vande00,recio06}.  
Recently, the possibility of
different helium abundances among stars within individual GCs has also been
suggested to explain not only the presence of multiple clumps and extended blue
tails on the horizontal branch, but also multiple parallel main sequences and
subgiant branches \citep[][and references therein]{dant08}. This working
hypothesis relies on the well established anti-correlations between the
molecular band-strengths of CN and CH \citep{smith87,Kraft94} and between O--Na
and Mg--Al measured in both evolved (red giant [RG] and horizontal branch [HB]) 
and unevolved (main sequence [MS]) stars of GCs investigated with high-resolution
spectroscopy \citep{pila83,rami02,grat04}.  Furthermore, deep Hubble Space
Telescope photometry has disclosed the presence of multiple stellar populations
in several massive GCs.  Together with the long-known case of $\omega$ Centauri
\citep{ande02,bedi04}, parallel stellar sequences have now been detected in a
number of GCs: \ngc{2808} \citep{dant05,piot07}, M54 \citep{sieg07} and \ngc{1851}
\citep{cala07,milo08}.  In some of these cases the multiple sequences might be
explained by sub-populations differing either in He abundance
\citep{norr04,dant05,dant08,piot07}, or in CNO abundance.

We still lack a direct detection of helium variations among the sub-populations
in a GC, because helium absorption lines in the visible spectral region appear only at 
effective temperatures hotter than 10,000~K \citep{behr03a, behr03b,moeh04b}.
Such high temperatures in a GC are typical of hot and extreme HB stars but,
unfortunately, their helium abundances cannot be adopted to constrain the original
helium content because gravitational settling and/or radiative levitation can affect
their surface abundances.  In a recent investigation \citet{vill09} identified
helium absorption lines in a few warm cluster HB stars that should be weakly
affected by these mechanisms, but the range in effective temperature remains
very narrow.  

Interestingly enough, \citet{pres09} has identified both HeI and HeII emission
and absorption lines in field RR Lyrae stars. The lines appear during the
rising branch of the light curve and appear to be the consequence of shocks
propagating through the extended atmosphere soon after the phase of minimum
radius \citep{bono94,chad96}. The idea is not new and dates back at least to
\citet{wall59}, who detected helium emission lines in his seminal investigation
of Type II Cepheids.  The key advantage of RR Lyrae stars, when compared with
similar detections of helium lines in hot and extreme HB stars \citep{behr03b}
is that they have an extensive convective envelope, and therefore they are
minimally affected by gravitational settling and/or radiative levitation
\citep{mich04}.  The drawback is that quantitative measurement of the helium abundance
requires hydrodynamical atmosphere models simultaneously accounting for
time-dependent convective transport and radiative transfer together with the
formation and the propagation of sonic shocks. 

A new approach has been suggested by \citet{dupr11}, who detected the
chromospheric HeI line at 10830$\,$\AA\ in a few RG stars having similar magnitudes
and colors in $\omega$~Cen.  The He line was detected in more metal-rich stars
and it seems to be correlated with Al and Na, but no clear correlation was found
with Fe abundance. However, a more detailed non-LTE analysis of the absolute
abundance of He is required before firm conclusions can be drawn concerning the
occurrence of a spread among the different sub-populations.

To overcome the thorny problems connected with the spectroscopic measurements,
\citet{iben68} suggested the use of the R parameter---the  ratio of the number
of HB stars to the number of RGB stars brighter than the flat 
part of the HB---to estimate the initial He abundance of cluster stars. Two other
independent constraints on the overall helium content in GCs were suggested by
\citet{capu83}. The $\Delta$ parameter, which is the difference in magnitude
between the MS at $(\bmv)_\circ=0.7$ and the luminosity level of the flat part of the
HB, and the A parameter, which is the mass-to-luminosity ratio of RR Lyrae
stars. The pros and cons of these parameters have been discussed in a thorough
investigation by \citet{sand00}. Theoretical and empirical limitations affecting the
precision of the R parameter have been also discussed by \citet{zocca00},
\citet{riel03}, \citet{cass03} and \citet{sala04}.

The key feature of all three parameters is that they are directly or indirectly
connected with the HB luminosity level. In spite of recent improvements in
photometric precision and the increased use of a multi-wavelength approach, we
still lack firm empirical methods to predict the HB luminosity level in GCs.
This problem is partially due to substantial change in the HB morphology 
when moving from metal-poor (generally blue HB) to metal-rich (red HB) GCs.
Additionally, we still lack a robust diagnostic to constrain the actual off-ZAHB
evolution of HB stars \citep{ferr99, dice10, cass11}.   
 
To overcome these longstanding problems we propose a new parameter---which we
call $\Delta \xi$---to estimate the helium content of GCs. The $\Delta
\xi$ parameter is the difference in magnitude between the RGB bump and a benchmark
defined as the point on the (essentially unevolved) MS at the same color as the RGB bump.  
In the following we discuss the theoretical framework adopted and the sample 
of globular clusters we have selected to calibrate this new parameter.   

%%%%%%%%%%%%%%%%%%%%%%%%%%%%%%%%%%%%%%%%%%%%%%%%%%%%%%%%%%%%%%%%%%%%%%%%%%%%%%%%%%%%%%%%%%%%%%%%%%%
\section{Theoretical framework}

To provide a quantitative theoretical framework for inferring the helium content of GCs 
we adopted the evolutionary models 
provided by \citet{piet04,piet06}\footnote{See the BaSTI web site at the following URL:
{\tt http://193.204.1.62/index.html}}. These models were computed using a recent version 
of the FRANEC evolutionary code \citep{c&s89}. The reader interested in a detailed 
discussion of the adopted input physics is referred to the above papers. Here we 
mention only that we adopted the radiative opacity tables\footnote{The reader interested in a 
detailed description is referred to the the following URL: http://opalopacity.llnl.gov/opal.html}  
from the Livermore group \citep{i&r96} for interior temperatures higher than $10^4$ K, 
and from \citet{ferg05} for the atmospheres. Thermal conduction is accounted 
for following the prescriptions by \citet{pote99}. We have updated the
plasma-neutrino energy loss rates according to the results provided by \citet{haft94}. 
The nuclear reaction 
rates have been updated using the NACRE data base \citep{angu99}, with the exception 
of the $^{12}C$($\alpha$;$\gamma$)$^{16}O$ reaction; for this reaction, we adopted the recent 
determination by \citet{kunz02}. We also adopted the equation of state (EOS) provided 
by A. Irwin\footnote{A detailed  description of this EOS can be found at the following URL: 
http://freeeos.sourceforge.net}. 

The adopted $\alpha$-enhanced mixture follows the prescriptions by \citet{sala97} 
and by \citet{s&w98}, with a Solar iron content of 0.0198 from \citet{g&n93}, 
a primordial helium content of Y=0.245 \citep{cass03}, and a helium-to-metal 
enrichment of $\Delta \hbox{\rm Y}/\Delta \hbox{\rm Z}$=1.4 \citep{piet04}.   
The cluster isochrones were transformed into the observational plane using the 
bolometric corrections (BCs) and the color-temperature relations (CTRs) provided 
by \citet{c&k03}. 

Note that together with the $\alpha$-enhanced models, we also considered cluster isochrones 
constructed assuming CNO-Na abundance anti-correlations \citep{piet09}. 
These sets of isochrones have the advantage of having been constructed for the same 
iron abundances as the $\alpha$-enhanced models. The heavy-element mixture adopted in 
these evolutionary models is defined as {\em extreme} \citep[see Table~1 in ][]{piet09},
and when compared with the canonical $\alpha$-enhanced mixture includes an increase in 
N and Na abundance of 1.8 and 0.8 dex and a decrease in C and O abundance of 0.6 
and 0.8 dex, respectively. We did not include the anti-correlation between Mg and Al, 
since its impact on evolutionary models is negligible \citep{sala06}.  
The transformation of these evolutionary models into the observational plane does
require a set of bolometric corrections (BCs) and color-temperature relations (CTRs) 
computed for the same mixtures; these became 
available only very recently \citep{sbo11}. However, \citet{piet09} found that 
BCs and CTRs computed for simple $\alpha$-enhanced mixtures mimic the same behavior. 
Moreover, \citet{dice10} found that BCs and CTRs hardly depend---at fixed total 
metallicity---on changes in the detailed mixture. The total metallicity of the atmosphere models 
adopted to transform the $\alpha$- and CNO-enhanced models was estimated using the 
\citet{sala93} relation with [$\alpha$/Fe]=0.62 dex. 

Finally, we mention that current isochrones together with similar predictions available in the 
literature \citep{vand00,dott07,bert08} give ages for GCs that are, within the errors, 
consistent with the WMAP age for the Universe of 13.7 Gyr \citep{koma11,lars11}, and 
provide a quite similar age ranking for the Galactic GCs \citep{mari09,cass11}.

%%%%%%%%%%%%%%%%%%%%%%%%%%%%%%%%%%%%%%%%%%%%%%%%%%%%%%%%%%%%%%%%%%%%%%%%%%%%%%%%%%%%%%%%%%%%%%%%%%%
\section{Optical data sets}

To provide a solid empirical basis for validating the $\Delta \xi$ parameter we
selected a sample of relatively luminous GCs (for good statistics), covering a
broad range in metallicity (--2.45$\le$[Fe/H]$\le$--0.70), with different
structural parameters, and all having relatively low foreground reddening
($E(\bmv) \leq 0.14$, except for \ngc{6352} with $E(\bmv)=0.21$~mag).
Moreover and even more importantly, we selected GCs for which precise and relatively 
deep photometry is available in the {\it BVI\/} photometric bands.  To reduce the 
possibility of any subtle errors caused by the
absolute photometric calibration or by the reduction strategy, we only adopted
catalogs available in the data base maintained by co-author PBS. We ended up
with a sample of almost two dozen GCs; their names, reddenings, metallicities
and total visual magnitudes are listed in the first four columns of Table~1.      
Note that the cluster iron abundances are based on the metallicity scale provided 
by \citet{ki03} based on FeII lines in high-resolution spectra. For the clusters 
for which the Kraft \& Ivans metallicity was not available we adopted the 
iron abundances on the metallicity scale by \citet{carr09} calibrated 
using FeI lines in high-resolution spectra. The mean difference between the 
two metallicity scales is minimal ($\lesssim$0.1 dex) and vanishingly small in the 
metal-poor regime (see their Fig.~10).    

All of these clusters have been calibrated to the Johnson-Cousins {\it BVI\/}
photometric system of \citet{land73,land92} as described by \citet{stet00,stet05}. 
In every case the corpus of observations is quite rich, ranging from a
minimum of 118 (\ngc{6362} in $B$) to a maximum of 3939 (\ngc{6341} in $I$) individual CCD images
per filter in each cluster (median:  946 images per filter per cluster). 
The instrumental observations have been
referred to the standard system by means of local standards within each field. 
The GCs in our sample have calibrated data in all three of the $B$, $V$, and $I$ bands
based on a minimum of 118 (\ngc{6362}) and a maximum of 3690 (\ngc{6341}) local standards
(median: 972).  Observations for each of the different clusters were obtained
during a minimum of 7 (\ngc{6541}) and a maximum of 44 (\ngc{6341}) different
observing runs (median: 16).  It must be stressed, however, that since our
proposed methodology relies on the difference in apparent magnitude between two
stellar sequences {\it at fixed color\/} absolute photometric zero points are
unnecessary, and possible errors in the color transformations of the photometric
calibration are irrelevant (provided any such errors are unaffected by stellar surface
gravity).  The principal observational requirement of this
work is that the photometry be consistently precise and {\it linear\/} over the
required magnitude range.  We believe that our data set, based on a very large
number of individual CCD images from many different observing runs and a
substantial and homogeneous body of local photometric standards for each cluster, guarantees
that this requirement is met to a higher degree than in any previous photometric surveys
of Milky Way GCs.  

%_______________________Figure 1  
\begin{figure*}
\includegraphics[height=0.75\textheight,width=0.95\textwidth]{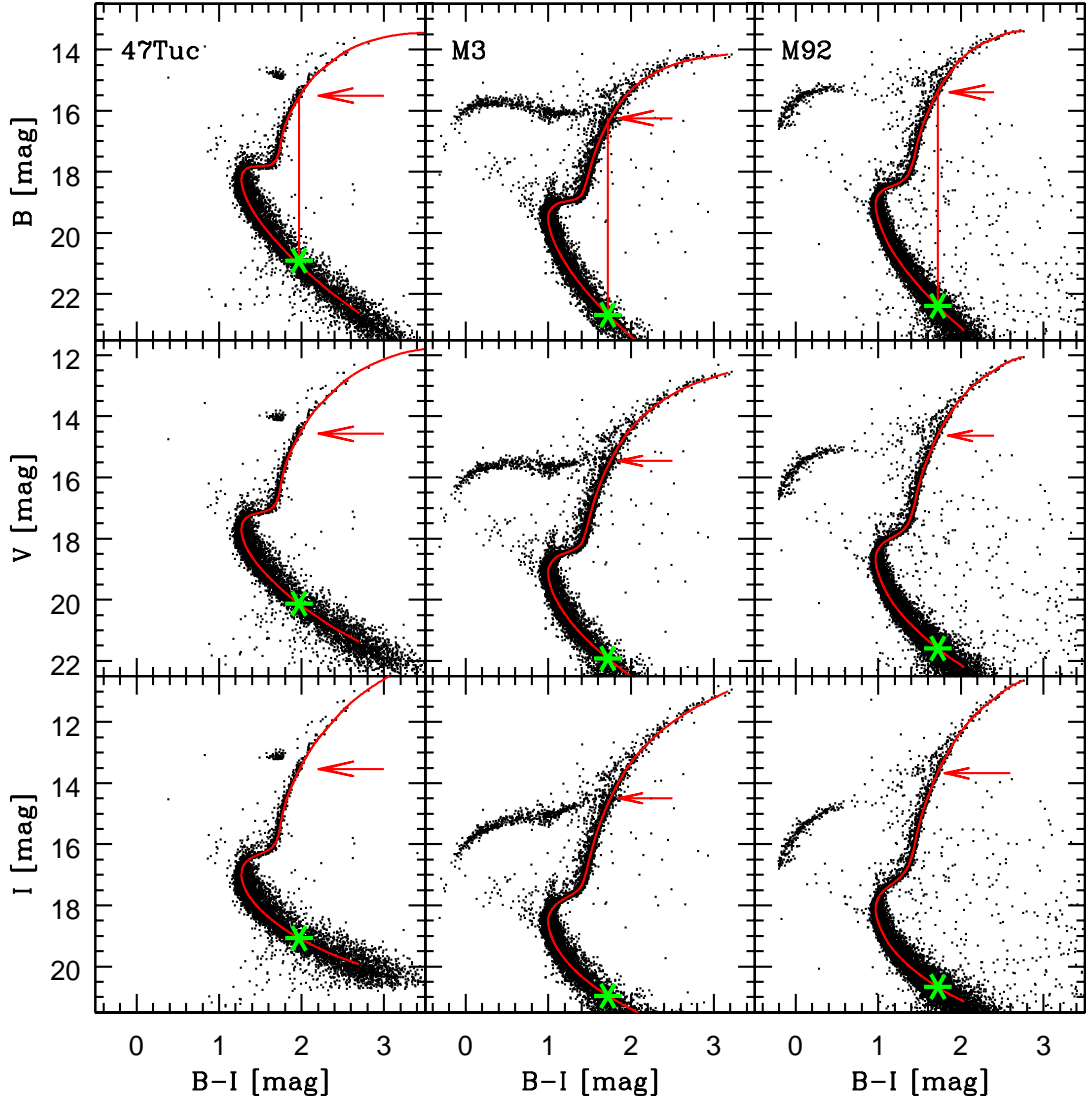}
\label{plotP}
\caption{Top -- $B$,\bmi\ color-magnitude diagram from left to right of a
metal-rich (\feh=--0.70, 47 Tuc), a metal-intermediate (\feh=--1.50, M3) and a
metal-poor (\feh=--2.38, M92) GC. The red solid curves show the cluster ridge
lines. The red arrows mark the position of the RGB bump, while the green asterisks 
show the MS benchmark. The vertical red lines show the difference in
magnitude between the RGB bump and the MS benchmark.  Middle -- Same as the top,
but in the $V$,\bmi\ color-magnitude diagram.  Bottom -- Same as the top, but in the
$I$,\bmi\ color-magnitude diagram.}
\end{figure*}

For each GC we estimated the ridge line in both the $V$,\bmv\ and the $V$,\vmi\
color-magnitude diagram (CMD).  The exact approach adopted to estimate the ridge 
lines will be described in a future paper (Ferraro et al.\ 2011, in preparation).
Fig.~1 shows three different CMDs for three clusters that are representative of
metal-rich (47 Tuc), metal-intermediate (M3) and metal-poor (M92) GCs. The red
solid curves display the ridge lines that we have estimated. The approach we
adopted to estimate the magnitude and the color of the RGB bump has already
been described in \citet{dice10}. The difference between the current $V$-band
RGB bumps and those estimated by \citet{dice10} is on average 0.00$\pm$0.04 mag
(12 GCs). The difference compared to the RGB bumps estimated by \citet{riel03} is
+0.01$\pm$0.07 mag (12 GCs), while the difference with the RGB bumps provided by
\citet{cass11} is  --0.03$\pm$0.05 mag. We performed a series of tests to
estimate the accuracy of the position of the bump along the RGB using different
colors (\bmi, \bmv, \vmi) and we found that among them \vmi\ provides the 
most precise measurements. We plan to provide a more detailed discussion 
concerning the pros and cons of both optical and near-infrared colors to estimate 
the $\Delta \xi$-parameter in a future paper. 
The $B,V$ magnitudes and the \vmi\ colors of the RGB bumps are
listed in columns 5, 6 and 7 of Table~1. The $B,V$ magnitudes of the MS benchmark  
having the same \vmi\ color as the RGB bump were estimated by interpolating the
ridge line (see the green asterisks plotted in Fig.~1) and they are listed in
columns 8 and 9 of Table~1.  

The data in Table~1 confirm the visual impression from Fig.~1 that the
luminosity of the RGB bump varies inversely with the metallicity of the cluster.  
The $\Delta \xi$ parameter is therefore
more prone to Poisson statistics and random photometric errors at the metal-poor end since in this
regime the bump falls in a region of the RGB where the lifetime is shorter than
at fainter magnitudes and the relative number of RG stars steadily decreases.
The color of the bump also becomes redder relative to the turnoff due to the slope of the RGB, and the
apparent magnitude of the MS benchmark becomes correspondingly fainter.

%%%%%%%%%%%%%%%%%%%%%%%%%%%%%%%%%%%%%%%%%%%%%%%%%%%%%%%%%%%%%%%%%%%%%%%%%%%%%%%%%%%%%%%%%%%%%%%%%%%%%%
\section{Comparison between theory and observations}

%_______________________Figure 2  
\begin{figure*}
\includegraphics[height=0.65\textheight,width=0.95\textwidth]{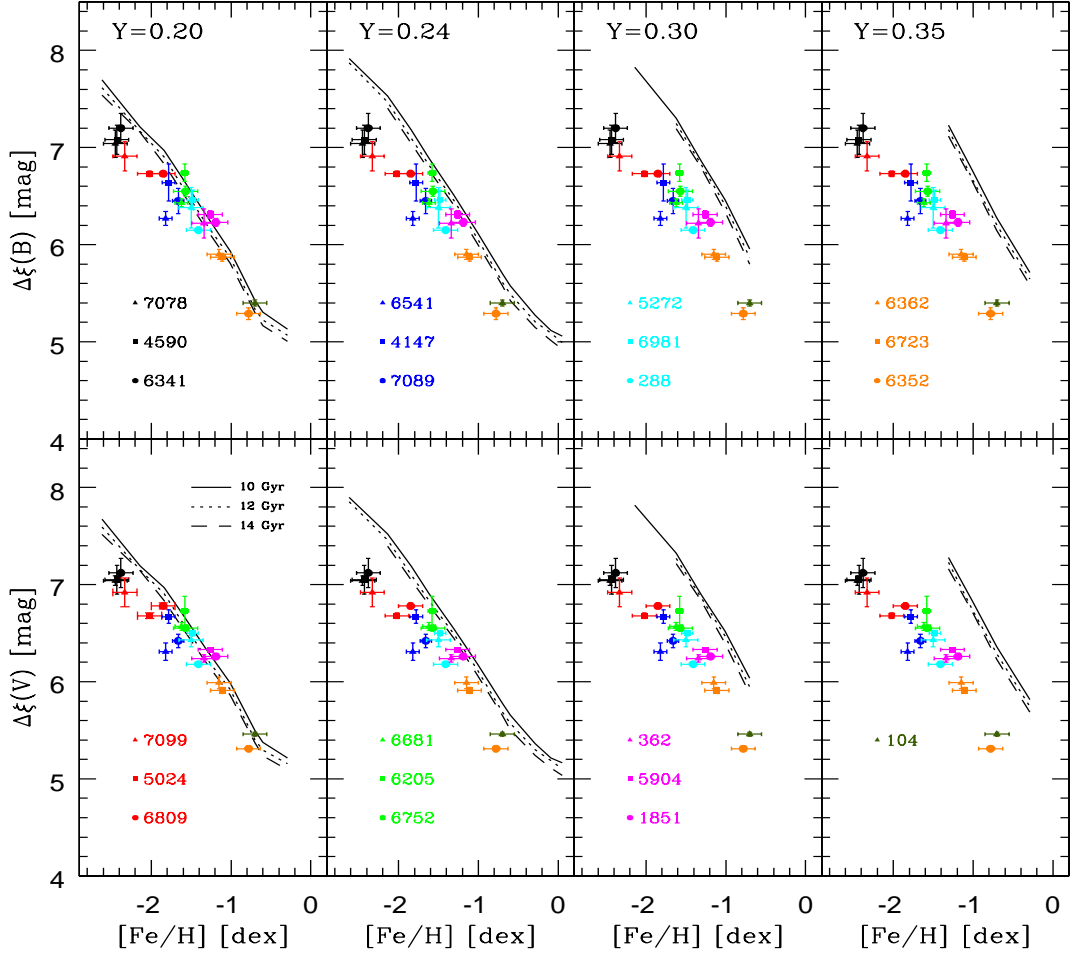}
\label{plotP}
\caption{Top -- Difference in the $B$-band magnitude between the RGB bump and the MS benchmark 
for the selected GCs as a function of the iron abundance. The selected GCs are 
marked with different symbols and colors. The dashed, dotted and solid lines display
theoretical predictions for three different cluster ages (10,12,14 Gyr) according
to the BASTI data base. The adopted cluster isochrones are based on evolutionary models
computed assuming an $\alpha$-enhanced mixture, different iron abundances
(--2.6$\lesssim$ \feh $\lesssim$ --0.3 dex) and five different He contents (Y=0.20--0.40). 
Predictions for Y=0.40 are not plotted, since the bump was identifiable only in a few 
isochrones.  Bottom -- Same as the top, but for the $V$-band magnitude.
}
\end{figure*}

We estimated the $\Delta \xi$ parameter in two different bands ($B$ and $V$);  these 
results are listed in columns 10 and 11 of Table~1, together with their intrinsic 
errors, and plotted in Fig.~2.  The adopted cluster isochrones were retrieved 
from the BASTI data base.  In particular, we estimated the predicted $\Delta \xi$ parameter 
using cluster isochrones based on evolutionary models constructed for iron abundances covering 
the entire range typical of old, low-mass stars, $\alpha$-enhanced ([$\alpha$/Fe]=+0.40 dex) mixtures 
and a mass-loss rate {\em \`a la\/} Reimers with $\eta$=0.4. We adopted five different helium 
abundances, namely Y=0.20, 0.245, 0.30, 0.35 and 0.40 (relative abundance in terms of mass 
fraction). The models with  Y=0.20, 0.30, 0.35 and 0.40 were constructed assuming a constant 
helium content over the entire metallicity range, while those with Y=0.245 were constructed 
assuming a helium-to-metal enrichment ratio \dydz=1.4 (see \S2). Note that the 
models with Y=0.20 have been specifically computed for this experiment, but they will become 
available at the web site mentioned above. To constrain the possible dependence on cluster age, 
we adopted three different values spanning the generally accepted range of GC
ages, namely 10, 12, 14 Gyr (see dashed, dotted and solid lines in Fig.~2).       

The predictions for Y=0.40 were not included in Fig.~2 because the bump was 
identifiable only in a few metal-rich compositions and for the older ages. The increase 
in helium causes, at fixed age and metal content, a decrease in the envelope opacity. 
This means that the maximum depth reached by the outer convective region during the 
first dredge-up becomes shallower and the discontinuity in the hydrogen profile left over 
by the retreating convective envelope is milder. Therefore, the H-burning shell 
crosses the chemical discontinuity at brighter magnitudes during faster evolutionary 
phases and, as a result, the bump becomes less and less evident.   

The main advantages in using the $\Delta \xi$ parameter to estimate the helium content 
are the following:
{\em i}) An increase in He content causes, at fixed cluster age and iron content,
a decrease in the envelope opacity and an increase in the mean molecular weight. This 
means that an increase in He makes the RGB bump systematically brighter and the MS 
benchmark systematically fainter. 
The sensitivity to helium is very good, and indeed the derivatives for 
$\alpha$-enhanced structures, at fixed iron and age, attain values similar 
to the $\Delta$ parameter (see Table~2).
{\em ii}) The sensitivity to age is small, and indeed the derivatives for 
$\alpha$-enhanced structures, at fixed iron and helium, attain minimal 
values.
{\em iii}) There is no dependence on photometric zero-point or interstellar
extinction or reddening.
{\em iv}) The derivative is almost linear over the entire metallicity 
range (see Fig.~2). This evidence is supported by both theory and observation. 
The R parameter, in contrast, shows a well defined change in the slope for 
[Fe/H]$\approx$--0.70 dex \citep{zocca00}.
{\em v}) There is no dependence on HB evolutionary effects. In contrast, the R, $\Delta$,
and A parameters are all affected by the off-ZAHB evolution of HB stars.

The use of the $\Delta \xi$ parameter has two main drawbacks:
{\em i}) Precise, linear photometry from the RGB bump down to the MS benchmark 
is required for photometric data sets large enough to properly identify the RGB bump.
{\em ii}) There is a definite metallicity dependence, and indeed the derivative
($\sim$1 mag/dex) is at least 0.7 dex larger than for the other helium indicators 
(see Table~2). The above points confirm that robust He estimates based on 
the $\Delta \xi$ parameter do require precise photometry and accurate 
spectroscopic iron abundances.

The positive features of the predicted $\Delta \xi$ parameter allow us to 
derive  an analytical relation with iron and helium content. Using a multi-linear 
regression over the entire set of $\Delta \xi^B$- and $\Delta \xi^V$ parameter, 
we found:\\      

\noindent
$Y=-0.525\pm0.037+0.158\pm0.007\,\Delta \xi^{B}+0.190\pm0.008\,\hbox{\rm [Fe/H]}\;\;\sigma=0.018$\\
\hspace*{16truecm}[1]\\
\noindent
$Y=-0.537\pm0.037+0.157\pm0.007\,\Delta \xi^{V}+0.180\pm0.008\,\hbox{\rm [Fe/H]}\;\;\sigma=0.018$\\

\noindent 
where $\sigma$ is the standard deviation and the other symbols have their usual meaning.
Note that we have neglected the age dependence because the dependence on this
parameter is imperceptible within the typical age range of GCs.  

The above analytical relations do allow us to provide individual helium estimates for GCs. 
However, the comparison between theory and observation indicates that the predicted 
$\Delta \xi$ parameters are systematically larger than observed and the discrepancy 
steadily increases when moving from lower to higher helium contents. Moreover, the 
discrepancy, at fixed helium content, increases when moving from metal-rich to metal-poor 
GCs. To represent the observed, empirical values of $\Delta \xi$ as a function of metallicity,
we performed linear least-squares fits to the data in Table~1 and obtained:\\

\noindent
$\Delta \xi^{B}= 4.75\pm0.13 - 1.04\pm0.08\,\hbox{\rm [Fe/H]} \;\;\sigma=0.08$~mag\\
\hspace*{16truecm}[2]\\
\noindent
$\Delta \xi^{V}= 4.85\pm0.12 - 1.00\pm0.07\,\hbox{\rm [Fe/H]} \;\;\sigma=0.09$~mag\\

The above fits were performed by using as weights the intrinsic errors on the
$\Delta\xi$ values. The anonymous referee noted that the slope of the $\Delta\xi$
relations might not be linear over the entire metallicity range. In particular, the
slope seems to become steeper in the metal-rich regime. However, this inference is
based on only two GCs (NGC~6352, NGC104).  Additional data in the high metallicity
tail of GGCs are required to reach more quantitative conclusions. However, to
constrain the impact of a possible change in the slope we performed the same fits
by using only GCs more metal-poor than \feh$\le$-1 dex. We ended up with a sample
of 20 GCs and found:\\

\noindent
$\Delta \xi^{B}= 5.00\pm0.15 - 0.90\pm0.09\,\hbox{\rm [Fe/H]} \;\;\sigma=0.13$~mag\\
\hspace*{16truecm}[3]\\
\noindent
$\Delta \xi^{V}= 5.18\pm0.13 - 0.81\pm0.07\,\hbox{\rm [Fe/H]} \;\;\sigma=0.17$~mag\\

The new fits are characterized by slopes that are slightly shallower and
zero-points that are mildly larger, but the two sets of least-squares solutions
agree within 1$\sigma$.

The small intrinsic dispersions of the above empirical relations suggest that the 
$\Delta \xi$ parameter can be adopted to identify GCs with peculiar He abundances. 
The $\Delta \xi$ values listed in Table~1 and the analytical relations based
on the less metal-rich GCs indicate that the possible dispersion in helium among
the target GCs is minimal. We found that only two GCs---M13 and NGC~6541---show
differences larger than 1.5$\sigma$ both in the $\Delta \xi^{B}$ and
$\Delta \xi^{V}$ parameters. However, the discrepancy concerning
NGC~6541 should be treated with caution, since it and NGC~6352 are
the two GCs in our sample with the largest reddening. We are only left with M13
for which the discrepancy ranges from 1.3 to 2.6$\sigma$ in the $V$ and
in the $B$ bands, respectively. By using the theoretical relations [1] we found
that the quoted differences imply on average a difference in helium content,
when compared with the bulk of GCs, that is of the order of 1$\sigma$
($\Delta Y$=0.031$\pm$0.025). The lack of a clear evidence of helium
enhancement in M13 soundly supports the recent results obtained by
\citet{sand10} using the R parameter. Unfortunately, we cannot extend the
current differential analysis concerning the helium content to more
metal-rich GCs, due to the scarcity in our sample of GCs in this
metallicity regime.

In this context it is also noteworthy that if the He content is {\it the\/} 
second parameter affecting the HB morphology \citep{suda06}, the $\Delta \xi$ 
parameter should be able to prove it.
At the present moment, the empirical evidence does not allow us to reach a firm 
conclusion. The classical couple of second parameter GCs M3 and M13 
\citep{john05}---i.e. two GCs that within the errors have the same chemical 
composition, but different HB morphologies--- show a mild difference 
($\approx$2$\sigma$) in $\Delta \xi$ ($\Delta \xi^{B}$= 6.38$\pm$0.21 [M3] vs 
6.74$\pm$0.09 [M13];  $\Delta \xi^{V}$= 6.43$\pm$0.07 [M3] vs 6.73$\pm$0.15 [M13]) 
level. On the other hand, the other classical couple of second parameter GCs 
NGC~288 and NGC~362 \citep{bell01} do not show, within the errors, any difference. 
The current scenario might be confused by systematic uncertainties affecting either 
the chemical composition (iron, $\alpha$-enhancement, CNO-enhancement) or the 
cluster age which are {\it not\/} directly correlated with a difference in He. 
Clearly, the simple-minded notion that He is {\it the\/} second parameter is 
not obviously correct; other factors, which are not immediately obvious, will 
have to be taken into account.     

The comparison between the empirical and predicted relations indicates that the discrepancy 
in the $B$ band for the lowest helium content (Y=0.20) is negligible in the metal-rich 
regime (\feh$\approx$--0.8 dex) and slightly smaller than 5$\sigma$ ($\Delta B=0.26$ mag) 
in the metal-poor regime (\feh$\approx$--2.2 dex). The difference in magnitude with canonical 
helium models ranges from $\Delta B=0.26$ (5$\sigma$, \feh$\approx$--0.8 dex) to $\Delta B=0.51$ 
(10$\sigma$, \feh$\approx$--2.2 dex) and becomes even larger for He-enhanced models 
(Y=0.30:  0.64, 0.89 mag; Y=0.35: 0.95, 1.21 mag in the metal-rich and in the metal-poor 
regime, respectively). The difference hardly depends on the adopted photometric band, 
and indeed the $V$-band difference with the lowest helium models is negligible 
in the metal-rich regime and equal to $\Delta V=0.20$ in the metal-poor regime 
(4$\sigma$, \feh$\approx$--2.2 dex). The difference with canonical helium models ranges 
from $\Delta V=0.25$ (5$\sigma$, \feh$\approx$--0.8 dex) to $\Delta V=0.45$ (9$\sigma$, 
\feh$\approx$--2.2 dex).

%_______________________Figure 3  
\begin{figure*}
\includegraphics[height=0.65\textheight,width=0.95\textwidth]{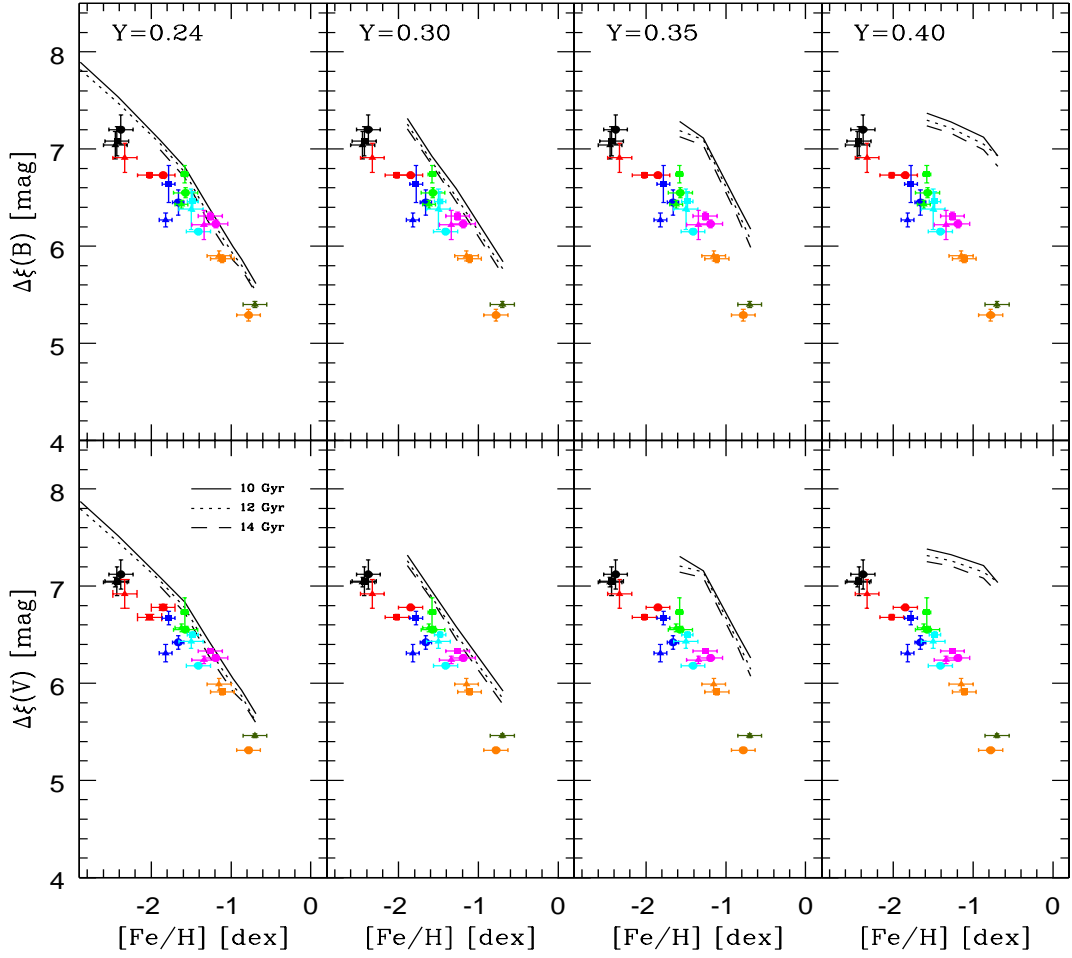}
\caption{Same as Fig.~2, but the stellar isochrones are based on evolutionary models
constructed assuming an  $\alpha$- and CNO-enhanced mixture. The adopted He abundances
range from Y=0.24 to Y=0.40.
}
\end{figure*}

The above results indicate either that the primordial helium content is roughly 20\% 
lower than estimated using the CMB \citep[WMAP,~][]{stei10} and the HII regions 
in metal-poor blue compact galaxies \citep{oliv04, izot10, peim10} 
or that theory overestimates the luminosity of the RGB bump.  
The latter working hypothesis is also supported by recent results by \citet{meis06}
and more recently by \citet{dice10} and by \citet{cass11}. The luminosity 
predicted by current evolutionary models is brighter than observed in actual GCs. Note 
that the current discrepancy is more striking than the previous estimates, since the 
$\Delta \xi$ parameter is minimally affected by uncertainties in distance, age, 
reddening, and HB luminosity level. 

To further constrain the nature of the possible culprits, we also tested the 
role played by possible overabundances in both $\alpha$- and CNO elements simultaneously. 
The comparison of the observations with predictions based on evolutionary models 
constructed assuming an {\em extreme} chemical mixture (see \S2) is shown in Fig.~3. 
Note that the comparison was performed at fixed iron content, since the iron 
abundance is found to be constant even in GCs showing well defined anti-correlations in the 
lighter metals. The data plotted in this figure indicate that the impact of 
$\alpha$- and CNO-enhanced mixture on the current discrepancy is small.

%%%%%%%%%%%%%%%%%%%%%%%%%%%%%%%%%%%%%%%%%%%%%%%%%%%%%%%%%%%%%%%%%%%%%%%%%%
\subsection{Possible stellar modeling venues}

Some crucial details of both micro- and macro-physics assumptions in the
modeling of low-mass stars are far from being settled.  Some of them may affect, 
directly or indirectly, the bump position in isochrones. \citet{cass11}
already discussed a number of them (their Sect.~4.1), either estimating 
their impact on the bump and on the turnoff position from their own calculations
or using earlier results available in the literature. Among these uncertainties
atomic diffusion and convective overshooting are the most interesting ones. 
Here we give an example of their possible influence on evolutionary 
tracks in the theoretical H-R diagram for a model with $M=0.8\,M_\odot$ and 
chemical composition Z=0.001, Y=0.250. The adopted  chemical 
composition is $\alpha$-element enhanced:  it corresponds 
to \feh= --1.56 dex. Such a star has a turn-off age of $\approx 13$~Gyr, 
and therefore its RGB is quite representative for the RGB of an isochrone 
of this age. The evolutionary tracks were constructed using the GARSTEC code 
\citep{w&s08}, and the evolutionary models were computed with the 
updated $^{14}\mathrm{N}(p,\gamma)^{15}\mathrm{O}$ reaction 
\citep{mart08}, which was shown \citep{weis05} to increase 
the bump luminosity by $\approx 0.06$~mag.
Note that the BASTI evolutionary tracks do not account for this 
reaction; its inclusion would increase the discrepancy between the predicted 
and the observed $\Delta \xi$ parameter by $\approx$0.15 mag.

%_______________________Figure 4  
\begin{figure} 
\includegraphics[height=0.35\textheight,width=0.50\textwidth]{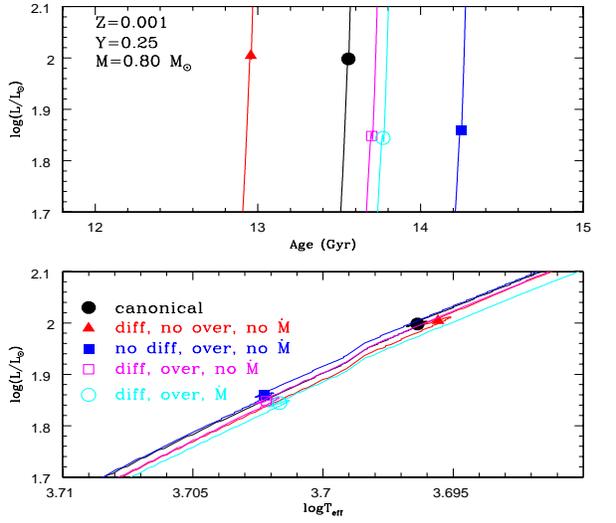}
\caption{Top -- Zoom of the evolutionary tracks around the position of the 
RGB bump for a model with $M=0.8\,M_\odot$ and chemical composition 
Z=0.001, Y=0.250, constructed assuming different assumptions concerning 
atomic diffusion, envelope overshooting and mass loss. The canonical 
evolutionary track (black line) neglects the above mechanisms. The position 
of the RGB bump is marked with different symbols. See text for more 
details.
Bottom -- Same as the top, but as a function of the logarithmic effective
temperature.} \label{}
\end{figure}

We start with a canonical model without convective overshooting and neglecting  
atomic diffusion (black line, top panel of Fig.~4)). 
In this case, the bump (which is not very pronounced for these stellar 
parameters) is located almost exactly at \lsun= 2.0. The inclusion of 
atomic diffusion (red line) decreases---as it is well-known---the 
main-sequence lifetime by $\approx$0.5~Gyr, but has only a minimal impact 
on the bump luminosity (triangle). Indeed, it becomes slightly brighter by only
$\approx$0.03 mag ($\approx$0.01~dex in $\log L$). We then included the 
overshooting, treated as a diffusive process as described in \citet{w&s08}, 
with a scale-length parameter of 0.04. The adopted value is approximately twice the 
value needed to fit the CMDs of open clusters \citep[see, e.g.~][]{magi10}.  Convective 
overshooting is applied at all the convective boundaries, and in particular 
at the lower boundary of the convective envelope. However, it is suppressed 
during the transient convective core phase of the pre-main sequence evolution.
The blue line plotted in the top panel of Fig.~4 shows that inclusion of the 
envelope overshooting makes the bump (filled blue square) fainter by 0.38~mag 
(0.15~dex in $\log L$) when compared with the canonical evolutionary track. 
This is more than enough to remove the discrepancy between the predicted 
and observed $\Delta \xi$ parameter. Note that the difference in luminosity 
between the bump in the canonical evolutionary track and in the track including 
overshooting is causing a minimal change in effective temperature 
($\Delta$ $T_{eff}\sim$65 K, see the bottom panel of Fig.~4), due to the steep
slope of the RGB along these evolutionary phases. This finding confirms the results 
originally suggested by \citet{alon93} concerning the dependence of the 
luminosity of the RGB bump on the envelope overshooting \citep[see also~][]{cass02,cass11}. 

As a further test, we included atomic diffusion together with envelope 
overshooting. The green line plotted in Fig.~4 shows that the bump (open 
green square) is getting even fainter, but the difference compared to 
the track including only the envelope overshooting is $\approx$0.05 mag 
(0.02~dex in $\log L$). Finally, together with envelope overshooting
and atomic diffusion we also included  mass loss according to the Reimers 
formula with the free parameter $\eta=0.6$ (cyan line). Note that this 
value is larger than the value typically adopted, $\eta=0.4$. However, 
the impact on the bump position (cyan filled circle) is negligible, 
the difference being of the order of a few hundredths of a magnitude. 

The inclusion of diffusion and/or convective overshooting has a minimal  
influence on the MS benchmark, since at the effective temperature of 
the bump, the corresponding mass on the main sequence is hardly affected 
(see \S 4.1). 
Therefore, the luminosity of the MS benchmark at fixed $T_\mathrm{eff}$ is 
unaffected.  This is the main difference with the results by \citet{cass11}, 
since along the MS they adopted the turn-off as reference point. 
Our indicator is age-independent to a high degree of confidence. 
However, there is an indirect effect: the convective overshooting moves 
the bump downwards along the RGB, which implies a blueward change of the 
bump $\log T_\mathrm{eff}$, in this case by $+0.0055$~dex. The MS benchmark has
therefore to move hotter (and brighter) on the MS by the same amount. Given 
the MS slope of $\triangle \log L / \triangle \log T_\mathrm{eff} = 7.5$, 
this implies a luminosity increase of $\approx$0.1 mag ($\approx$10\% in luminosity), 
thus decreasing the discrepancy between theory and observation 
even more. In summary, our test calculations show that reasonable assumptions 
concerning the convective overshooting from the bottom of the convective envelope 
goes in the right direction to significantly reduce the discrepancy, with the 
atomic diffusion playing a secondary role.

It is worth noting that the quoted numerical experiments were performed at 
fixed chemical composition and cluster age. This means that the possible 
explanation we are suggesting to remove the discrepancy between predicted and 
observed $\Delta \xi$ parameters needs to be tested, together with the new 
$^{14}\mathrm{N}(p,\gamma)^{15}\mathrm{O}$ reaction, in both the metal-poor 
and the metal-rich regime (Weiss et al.\ in preparation). 

Finally, it is worth mentioning that \citet{char07}, to explain the 
abundance anomalies of RGB stars, suggested the occurrence of a thermohaline 
instability along the RGB when the hydrogen burning shell reaches the layers 
with uniform chemical composition left behind by the first dredge-up. 
However, the occurrence of this phenomenon has a minimal impact on the 
luminosity of the RGB bump, since the hydrogen burning shell reaches 
the chemical discontinuity at earlier evolutionary phases. Moreover, 
recent evolutionary calculations, including state-of-the-art composition 
transport predictions, indicate that the thermohaline region does not 
reach the outer convective envelope \citep{wac11}. This limits the 
possibility for a non-canonical mixing process and for relevant changes 
in the chemical stratification above the hydrogen burning shell.

%%%%%%%%%%%%%%%%%%%%%%%%%%%%%%%%%%%%%%%%%%%%%%%%%%%%%%%%%%%%%%%%%%%%%%%%%%%%%%%%%%%%%%%%%%%%%%%%%%%%%%%%%%%%%%%
\section{Conclusions and final remarks}

We introduce a new $\Delta \xi$ parameter---the difference in magnitude between the 
RGB bump and the MS benchmark (defined as the point on the MS with the same color as the 
RGB bump)---to estimate the helium content of globular clusters. The advantages of the 
$\Delta \xi$ parameter are the following:
{\em i}) An increase in He content causes, at fixed age and iron content,
an RGB bump systematically brighter and an MS benchmark systematically fainter.  
The sensitivity to He is very good, and indeed the 
derivatives for $\alpha$-enhanced structures, at fixed iron and age, attain 
values (6.47--6.56 mag) similar to the $\Delta$ parameter (6.63 mag).
{\em ii}) It has a minimal sensitivity to age, and indeed the derivatives for
$\alpha$-enhanced structures, at fixed iron and helium, are minimal 
(--0.03 mag/Gyr).
{\em iii}) It has no dependence on uncertainties in the photometric zero-point or
the amount of reddening and extinction.
{\em iv}) Its sensitivity to helium is almost linear at every metal abundance.
{\em vi}) It is minimally affected by evolutionary effects. In contrast, the R, 
$\Delta$, and A parameters are all affected by the post-ZAHB evolution 
of HB stars.

The $\Delta \xi$ parameter has two main drawbacks:
{\em i}) Precise linear photometry from the RGB bump down to the MS benchmark is 
required, with large enough photometric data sets to properly characterize the RGB bump.
{\em ii}) The slope of the $\Delta \xi$ versus Y relation has a strong
metallicity dependence. The derivative is $\sim$--1 mag/dex/dex, i.e., a factor of
0.7 larger than the $\Delta$ parameter and $\approx$1 dex larger than the R 
and A parameters.  

In passing we note that the $\Delta \xi$ parameter, at constant helium-to-metal enrichment 
ratio, can also be adopted as a metallicity indicator. The slope of the metallicity 
dependence is similar to the slope of the RGB bump (1 vs 0.9), but the $\Delta \xi$ 
parameter is minimally affected by uncertainties in cluster distance and in reddening.  

We have measured the $\Delta \xi$ parameter in almost two dozen GGCs. The sample was
selected according to the following criteria: homogeneous photometry; low
reddening; a broad metallicity range (--2.5$\ltsim$ \feh $\ltsim$--0.7 dex); and a range of
structural properties.  We found that, at first glance, theory and observation
agree quite well if we assume a primordial He content of Y=0.20, and indeed they
agree reasonably well in both the metal-rich and the metal-intermediate regime.
However, the observed values in the metal-poor regime are 5$\sigma$ ($\Delta$B=0.26 mag) 
smaller than predicted. 
The difference with models assuming a more canonical primordial He abundance 
is larger, and ranges from  5$\sigma$ ($\Delta B=0.26$~mag) in the metal-rich 
regime to  4$\sigma$ ($\Delta B=0.51$~mag) in the metal-poor regime.  The difference 
becomes even larger when moving to higher He contents (such as are assumed in some 
current interpretations of the multiple populations found in some GCs).  
These results support previous evidence that current evolutionary prescriptions
generally overestimate the luminosity of the RGB bump \citep{meis06,dice10,cass11}.  

We computed a series of alternative evolutionary tracks at fixed age ($t=13$~Gyr,
M=0.80 M$_\odot$) and chemical composition (Z=0.001, Y=0.250) and we found that
plausible assumptions concerning the treatment of the envelope overshooting can
account for the discrepancy between theory and observations.  The atomic
diffusion and the mass loss may also play secondary roles. For a thorough discussion
of the possible impact of other physical parameters the reader is referred to
\citet{cass11}.  

The current debate concerning the primordial helium content is far from being settled. 
The primordial helium abundance mainly affects the third and the fourth acoustic peak 
of the CMB angular power spectrum. The seven years of observations by WMAP have provided 
an opportunity to improve the accuracy of the third peak. This means that He can be 
treated, for the first time, as a fitted parameter in the $\Lambda$CDM model. 
In particular, using a flat prior with 0.01$<$ $Y_P$ $<$0.8, \citet{lars11} found
$Y_P$=0.28$^{+0.14}_{-0.15}$ (see their Table~8).  This confirms the existence
of pre-stellar helium at least at a 2$\sigma$ level.  These findings were
strongly supported by \citet{koma11} who combined the seven years of WMAP data
with data from small-scale CMB experiments (ACBAR, \citealt{reic09}; QUaD, 
\citealt{brow09}).  They found a 3$\sigma$ detection of primordial He
($Y_P$=0.326$\pm$0.075 [68\% CL], see their figures 10 and 11). To further
constrain the primordial helium they adopted a very conservative upper limit
($Y_P$$<$0.3) and they found (0.23 $<$$Y_P$$<$0.3  68\% CL).  These geometric
determinations of the primordial helium abundance do not by themselves rule out an
initial Y value as small as 0.20.

According to the most recent calculations of state-of-the-art Big Bang Nucleosynthesis (SBBN) 
models the primordial helium content is Y$_P \approx 0.2483 \pm 0.0005$. The interested reader 
is referred to the comprehensive review by \citet{stei07}. This value agrees quite well with 
helium abundance estimates based on metal-poor extragalactic HII regions, namely 
Y${}_P \approx 0.240 \pm 0.006$, but the impact on these measurements of possible systematic 
errors is still controversial \citep{izot04, oliv04, fuku06, peim07}.    

In light of the above findings concerning the primordial He content, we followed this
path:  we estimated the primordial helium content using the analytical relations 
given in \S3 and the $\Delta \xi$ parameters listed in Table~1. We found 
Y${}_P =0.181 \pm 0.014$ ($\sigma$=0.034, $\Delta \xi^B$) and $0.183 \pm 0.014$
($\sigma$=0.033, $\Delta \xi^V$). 
The current Y${}_P$ values were estimated as weighted means, the errors are the standard error
of the mean
and account for the uncertainties in the coefficients of the analytical relations, 
while the $\sigma$'s are the standard deviations.  
It is worth noting that the Y${}_P$ values we found are upper limits to the primordial 
helium content, since some extra helium is dredged up to the surface by the deep 
penetration of the convective envelope during the so-called first dredge-up phase. 
However, the amount of extra helium is very limited, and indeed at fixed cluster age 
(12 Gyr) it ranges from 0.008 (\feh=--2.62 dex, M[TO]=0.80 M$_\odot$) to 0.02 
(\feh=--0.70 dex, M[TO]=0.90 M$_\odot$).    

The above estimates are within 1$\sigma$ of the estimates provided by the CMB experiments,
and by the analysis of lower main sequence stars \citep{casa07}, but are 2$\sigma$ from 
the estimates provided by the SBBN models and by the measurements of extragalactic HII 
regions. 

We also assumed that the current evolutionary models, when moving from the metal-poor 
to the metal-rich regime, on average underestimate the luminosity of the $\Delta \xi$ 
parameter by 0.35 mag because of incomplete input physics. Therefore, we 
can apply---as a preliminary correction---a systematic shift of 0.35 mag to the analytical 
relations. Note that the coefficients of the $\Delta \xi$ parameter are of the order of 0.016, 
so the current shift causes an increase in helium of a few percent.  Under these assumptions
we found a primordial helium content of Y${}_P$=0.236$\pm$0.014 ($\Delta \xi^B$) and 
0.238$\pm$0.014 ($\Delta \xi^V$). 

Consideration of the above results indicates that we still lack a convincing determination 
of the primordial He content based on stellar observables. However, if we trust the input 
physics currently adopted in evolutionary models, the existence of pre-stellar helium is 
established at least at the 5$\sigma$ level and increases to at least a
6$\sigma$ level, if we assume that current models underestimate the luminosity
of the RGB bump by a few tenths of a magnitude.  

No doubt the new and accurate CMB measurements by PLANCK \citep{ichi08} will shed new light 
on the long-standing problem of the abundance of the second most abundant element in the 
Universe. The stellar path seems also very promising, but we still need to pave it and GAIA 
is a fundamental step in this direction.

\acknowledgments
It is a real pleasure to thank S. Cassisi for many useful discussions concerning the 
evolutionary properties of low-mass stars and L. Althaus for detailed discussion 
concerning the thermohaline instability.  
We also acknowledge an anonymous referee for his/her positive comments on the 
results of this investigation and for his/her constructive suggestions that 
improved the content and the readability of the paper.
This project was partially supported by the PRIN-INAF2010 (PI: R. Gratton). One of 
us (G.B.) acknowledges support from the ESO Visitor Program.  
This publication makes use of data products from VizieR (Ochsenbein et al.\ 2000) and 
from the Two Micron All Sky Survey, which is a joint project of the University of 
Massachusetts and the Infrared Processing and Analysis Center/California Institute 
of Technology, funded by the National Aeronautics and Space Administration and 
the National Science Foundation. 

%%%%%%%%%%%%%%%%%%%%%%%%%%%%%%%%%%%%%%%%%%%%%%%%%%%%%%%%%%%%%%%%%%%%%%%%%%%%%%%%%%%%%%%

%%%%%%%%%%%%%%%%%%%%%%%%%%%%%%%%%%%%%%%%%%%%%%%%%%%%%%%%%%%%%%%%%%%%%%%%%%%%%%%%%%%%%
%                       tab 1 
%%%%%%%%%%%%%%%%%%%%%%%%%%%%%%%%%%%%%%%%%%%%%%%%%%%%%%%%%%%%%%%%%%%%%%%%%%%%%%%%%%%%%
\begin{centering}
\begin{deluxetable}{lcccccccccccc}
\rotate
\tabletypesize{\scriptsize}
\tablewidth{0pt}
\tablecaption{Apparent $B$- and $V$-band magnitude of the RGB bump and MS benchmark, 
together with the $(V-I)$ color of the bump. The reddening, the mean metallicity 
and the total visual magnitude for each cluster are also listed.}
\tablehead{
\colhead{ID (alias)}&
\colhead{E(B-V)\tablenotemark{a}}&
\colhead{[Fe/H]}&
\colhead{$M_{V}\tablenotemark{b}$}&
\colhead{$B_{bump}$}&
\colhead{$V_{bump}$}&
\colhead{(V-I)$_{bump}$}&
\colhead{$B_{MSB}$}&
\colhead{$V_{MSB}$}&
\colhead{$\Delta\xi^{B}$}&
 \colhead{$\Delta\xi^{V}$}\\
               &   mag  &   dex      & mag                     &  mag          &    mag          &    mag        &   mag          &    mag    &     mag     &      mag
}
\startdata
\ngc{7078} (M15) & $0.10$ & $-2.45\pm0.15\tablenotemark{c}$ & $-9.03$ & $16.07\pm0.01$ & $15.25\pm0.02$ & $1.02\pm0.04$ & $23.11\pm0.14$ & $22.29\pm0.05$ & $7.04\pm0.14$ & $7.04\pm0.05$\\
\ngc{4590} (M68) & $0.05$ & $-2.43\pm0.15\tablenotemark{c}$ & $-7.37$ & $15.95\pm0.02$ & $15.15\pm0.01$ & $1.01\pm0.01$ & $23.03\pm0.10$ & $22.20\pm0.07$ & $7.08\pm0.10$ & $7.05\pm0.07$\\
\ngc{6341} (M92) & $0.02$ & $-2.38\pm0.15\tablenotemark{c}$ & $-8.21$ & $15.40\pm0.01$ & $14.64\pm0.02$ & $0.96\pm0.02$ & $22.60\pm0.10$ & $21.76\pm0.08$ & $7.20\pm0.10$ & $7.12\pm0.08$\\
\ngc{7099} (M30) & $0.03$ & $-2.33\pm0.15\tablenotemark{c}$ & $-7.45$ & $15.52\pm0.01$ & $14.71\pm0.01$ & $0.99\pm0.02$ & $22.43\pm0.09$ & $21.63\pm0.05$ & $6.91\pm0.09$ & $6.92\pm0.05$\\
\ngc{5024} (M53) & $0.02$ & $-2.02\pm0.15\tablenotemark{c}$ & $-8.71$ & $17.36\pm0.02$ & $16.61\pm0.03$ & $0.95\pm0.18$ & $24.09\pm0.22$ & $23.29\pm0.18$ & $6.73\pm0.22$ & $6.68\pm0.18$\\
\ngc{6809} (M55) & $0.08$ & $-1.85\pm0.15\tablenotemark{c}$ & $-7.57$ & $15.12\pm0.01$ & $14.17\pm0.01$ & $1.09\pm0.01$ & $21.85\pm0.03$ & $20.95\pm0.03$ & $6.73\pm0.03$ & $6.78\pm0.03$\\
\ngc{6541}       & $0.14$ & $-1.82\pm0.08\tablenotemark{d}$ & $-8.52$ & $15.91\pm0.01$ & $15.00\pm0.01$ & $1.08\pm0.01$ & $22.18\pm0.07$ & $21.31\pm0.09$ & $6.27\pm0.07$ & $6.31\pm0.09$\\
\ngc{4147}       & $0.02$ & $-1.78\pm0.08\tablenotemark{d}$ & $-6.17$ & $17.48\pm0.01$ & $16.64\pm0.01$ & $0.97\pm0.01$ & $24.12\pm0.19$ & $23.31\pm0.07$ & $6.64\pm0.19$ & $6.67\pm0.07$\\
\ngc{7089} (M2)  & $0.06$ & $-1.66\pm0.07\tablenotemark{d}$ & $-9.03$ & $16.65\pm0.01$ & $15.83\pm0.01$ & $0.99\pm0.03$ & $23.10\pm0.13$ & $22.25\pm0.07$ & $6.45\pm0.13$ & $6.42\pm0.07$\\
\ngc{6681} (M70) & $0.07$ & $-1.62\pm0.08\tablenotemark{d}$ & $-7.12$ & $16.61\pm0.01$ & $15.58\pm0.01$ & $1.10\pm0.02$ & $23.04\pm0.04$ & $22.15\pm0.04$ & $6.43\pm0.04$ & $6.57\pm0.04$\\
\ngc{6205} (M13) & $0.02$ & $-1.58\pm0.05\tablenotemark{d}$ & $-8.55$ & $15.53\pm0.01$ & $14.72\pm0.01$ & $0.96\pm0.01$ & $22.27\pm0.09$ & $21.45\pm0.15$ & $6.74\pm0.09$ & $6.73\pm0.15$\\
\ngc{6752}       & $0.04$ & $-1.57\pm0.15\tablenotemark{c}$ & $-7.73$ & $14.49\pm0.01$ & $13.65\pm0.01$ & $1.02\pm0.01$ & $21.04\pm0.05$ & $20.20\pm0.03$ & $6.55\pm0.05$ & $6.55\pm0.03$\\
\ngc{5272} (M3)  & $0.01$ & $-1.50\pm0.15\tablenotemark{c}$ & $-8.88$ & $16.25\pm0.04$ & $15.45\pm0.02$ & $0.96\pm0.03$ & $22.63\pm0.21$ & $21.88\pm0.07$ & $6.38\pm0.21$ & $6.43\pm0.07$\\
\ngc{6981} (M72) & $0.05$ & $-1.48\pm0.07\tablenotemark{d}$ & $-7.04$ & $17.57\pm0.01$ & $16.70\pm0.01$ & $1.00\pm0.01$ & $24.03\pm0.10$ & $23.20\pm0.03$ & $6.46\pm0.10$ & $6.50\pm0.03$\\
\ngc{288}        & $0.03$ & $-1.41\pm0.15\tablenotemark{c}$ & $-6.75$ & $16.33\pm0.01$ & $15.48\pm0.01$ & $0.96\pm0.01$ & $22.48\pm0.03$ & $21.66\pm0.02$ & $6.15\pm0.03$ & $6.18\pm0.02$\\
\ngc{362}        & $0.05$ & $-1.34\pm0.15\tablenotemark{c}$ & $-8.43$ & $16.25\pm0.02$ & $15.38\pm0.01$ & $0.98\pm0.01$ & $22.47\pm0.15$ & $21.62\pm0.03$ & $6.22\pm0.15$ & $6.24\pm0.04$\\
\ngc{5904} (M5)  & $0.03$ & $-1.26\pm0.15\tablenotemark{c}$ & $-8.81$ & $15.82\pm0.01$ & $14.99\pm0.01$ & $0.99\pm0.01$ & $22.13\pm0.04$ & $21.32\pm0.02$ & $6.31\pm0.04$ & $6.33\pm0.02$\\
\ngc{1851}       & $0.02$ & $-1.19\pm0.15\tablenotemark{c}$ & $-8.33$ & $17.01\pm0.01$ & $16.13\pm0.01$ & $0.99\pm0.01$ & $23.24\pm0.04$ & $22.39\pm0.03$ & $6.23\pm0.04$ & $6.26\pm0.03$\\
\ngc{6362}       & $0.09$ & $-1.15\pm0.15\tablenotemark{c}$ & $-6.95$ & $16.49\pm0.01$ & $15.49\pm0.01$ & $1.06\pm0.01$ & $22.39\pm0.05$ & $21.48\pm0.06$ & $5.90\pm0.05$ & $5.99\pm0.06$\\
\ngc{6723}       & $0.05$ & $-1.11\pm0.15\tablenotemark{c}$ & $-7.83$ & $16.48\pm0.01$ & $15.58\pm0.01$ & $1.05\pm0.01$ & $22.35\pm0.04$ & $21.49\pm0.03$ & $5.87\pm0.04$ & $5.91\pm 0.03$\\
\ngc{6352}       & $0.21$ & $-0.78\pm0.15\tablenotemark{c}$ & $-6.47$ & $16.90\pm0.01$ & $15.75\pm0.01$ & $1.31\pm0.01$ & $22.19\pm0.06$ & $21.06\pm0.03$ & $5.29\pm0.06$ & $5.31\pm0.03$\\
\ngc{104} (47Tuc)& $0.04$ & $-0.70\pm0.15\tablenotemark{c}$ & $-9.42$ & $15.51\pm0.01$ & $14.57\pm0.01$ & $1.03\pm0.01$ & $20.91\pm0.03$ & $20.03\pm0.02$ & $5.40\pm0.03$ & $5.46\pm0.02$\\
\enddata
\tablenotetext{a}{Cluster reddening according to \citet{harr10}}
\tablenotetext{b}{Cluster absolute visual magnitude according to \citet{harr10}}
\tablenotetext{c}{Cluster iron abundance according to \citet{ki03,ki04}}
\tablenotetext{d}{Cluster iron abundance according to \citet{carr09}}
\end{deluxetable}
\end{centering}

%%%%%%%%%%%%%%%%%%%%%%%%%%%%%%%%%%%%%%%%%%%%%%%%%%%%%%%%%%%%%%%%%%%%%%%%%%%%%%%%%%%%%
%                       tab 2
%%%%%%%%%%%%%%%%%%%%%%%%%%%%%%%%%%%%%%%%%%%%%%%%%%%%%%%%%%%%%%%%%%%%%%%%%%%%%%%%%%%%%
\begin{deluxetable}{lccccccc}
\tabletypesize{\scriptsize}
\tablewidth{0pt}
\tablecaption{Derivative as a function of iron, helium and cluster age for the most
popular parameters adopted to estimate the helium content in GCs.}
\tablehead{
\colhead{Derivative}&
\colhead{$\Delta \xi^B_\alpha$}&
\colhead{$\Delta \xi^V_\alpha$}&
\colhead{$\Delta \xi^B_{\alpha, CNO}$}&
\colhead{$\Delta \xi^V_{\alpha, CNO}$}&
\colhead{$R$\tablenotemark{d}}&
\colhead{$\Delta$\tablenotemark{d}}&
\colhead{$A$\tablenotemark{d}}
}
\startdata
$\left(\frac{\partial }{\partial Y}\right)^{a}_{[Fe/H]=-1.3, t=12}$   & $6.47$  &$6.56$  & $5.07$  & $5.11$ & 11.43 &  6.63 & 1.50 \\
$\left(\frac{\partial }{\partial [Fe/H]}\right)^{b}_{Y=0.246, t=12}$  & $-1.08$ &$-1.03$ & $-1.04$ & $-1.00$& -0.08 & -0.37 & -0.05 \\
$\left(\frac{\partial }{\partial t}\right)^{c}_{[Fe/H]=-1.3, Y=0.246}$& $-0.03$ &$-0.03$ & $-0.04$ & $-0.04$& \ldots& \ldots&\ldots \\
\enddata
\tablenotetext{a}{The derivative (mag) was computed at fixed iron content and cluster ages over the entire range of He abundances covered by $\alpha$- and CNO-enhanced models (BASTI data base).}
\tablenotetext{b}{The derivative (mag/dex/dex) was computed at fixed helium content and cluster ages over the entire range of iron abundances covered by $\alpha$- and
CNO-enhanced models (BASTI data base).}
\tablenotetext{c}{The derivative (mag/Gyr) was computed at fixed helium and iron abundances over isochrones with ages ranging from 10 to 14 Gyr (BASTIatabase).}
\tablenotetext{d}{The derivatives were computed adopting $\alpha$-enhanced models (BASTI data base).}
\end{deluxetable}

\end{document}